\documentclass[acmsmall,screen,review]{acmart}
\renewcommand\footnotetextcopyrightpermission[1]{} 
\AtBeginDocument{%
  }
    
\usepackage{algorithm}
\usepackage{algorithmic}

\usepackage{utfsym}

\usepackage{tabularx} 
\usepackage{array}

\usepackage{multirow} 
\usepackage{booktabs} 
\usepackage{makecell} 
\usepackage{array} 
\usepackage{multirow} 

\usepackage{ragged2e} 

\usepackage{caption}
\usepackage{array} 
\usepackage{cellspace} 
\usepackage{array} 
\usepackage{cellspace} 
\usepackage{colortbl} 
\definecolor{lightgray}{gray}{0.8} 

\usepackage[utf8]{inputenc}
\usepackage{listings}
\usepackage{xcolor}

\lstdefinelanguage{Solidity}{
    keywords={contract, function, public, payable, mapping, address, uint256, msg, sender}, 
    keywordstyle=\color{cyan!80!black}\bfseries, 
    morestring=[b]", 
    stringstyle=\color{orange!80!black}, 
    morecomment=[l]{//}, 
    morecomment=[s]{/*}{*/}, 
    commentstyle=\color{gray!60}\itshape, 
    sensitive=true 
}

\begin{document}
\title{SoK: Security Analysis of Blockchain-based Cryptocurrency}

\author{Zekai Liu}
\affiliation{%
  \institution{Hainan University}
  \city{Haikou}
  \country{China}}
\email{lyxxxx1988@gmail.com}

\author{Xiaoqi Li}
\affiliation{%
  \institution{Hainan University}
  \city{Haikou}
  \country{China}}
\email{csxqli@ieee.org}

\renewcommand{\shortauthors}{Liu and Li}

\begin{abstract}
Cryptocurrency is a novel exploration of a form of currency that proposes a decentralized electronic payment scheme based on blockchain technology and cryptographic theory. While cryptocurrency has the security characteristics of being distributed and tamper-proof, increasing market demand has led to a rise in malicious transactions and attacks, thereby exposing cryptocurrency to vulnerabilities, privacy issues, and security threats. Particularly concerning are the emerging types of attacks and threats, which have made securing cryptocurrency increasingly urgent. Therefore, this paper classifies existing cryptocurrency security threats and attacks into five fundamental categories based on the blockchain infrastructure and analyzes in detail the vulnerability principles exploited by each type of threat and attack. Additionally, the paper examines the attackers' logic and methods and successfully reproduces the vulnerabilities. Furthermore, the author summarizes the existing detection and defense solutions and evaluates them, all of which provide important references for ensuring the security of cryptocurrency. Finally, the paper discusses the future development trends of cryptocurrency, as well as the public challenges it may face.
\end{abstract}


\maketitle
\fancyfoot{}
\pagestyle{plain} 

\section{Introduction}

In recent years, cryptocurrency flourish at an unprecedented rate \cite{catherine2024metaverse}. Their decentralized and anonymous characteristics draw widespread global attention. However, with the rapid expansion of the cryptocurrency market, security threats and attacks also present an increasingly severe situation \cite{weichbroth2023security}. Although blockchain technology, as the underlying support of cryptocurrency, provides a certain degree of security, its inherent technical complexity and emerging characteristics make security vulnerabilities difficult to completely avoid. Moreover, potential defects in the anonymity and traceability of cryptocurrency provide opportunities for malicious attackers.

Since the concept of cryptocurrency emerged, a series of major security incidents continuously occur, which not only cause severe damage to the blockchain ecosystem but also bring significant economic losses to individuals and institutions \cite{li2024guardians, li2021hybrid}. For example, the 2016 The DAO attack exposes the enormous risks of smart contract vulnerabilities, resulting in the theft of over \$50 million in cryptocurrency \cite{scharfman2024decentralized}; the 2017 Parity wallet smart contract vulnerability freezes over \$30 million in Ether \cite{bukhari2024secure}; the 2018 Coincheck trading platform theft results in losses of over \$500 million in NEM \cite{krause20251}; and the 2021 Poly Network cross-chain attack allows hackers to steal approximately \$610 million in cryptocurrency \cite{wu2024safeguarding}. These incidents fully demonstrate that cryptocurrency security issues become a critical factor restricting its healthy development.

Cryptocurrency face numerous security challenges, including double-spending attacks, Sybil attacks, denial-of-service (DoS) attacks, and selfish mining attacks, which urgently require solutions. Numerous scholars conduct extensive and in-depth research on these issues. For instance, Chen et al. (2022) comprehensively review the types of attacks on blockchain systems, defense mechanisms, and privacy-preserving techniques, emphasizing the critical need to balance security and performance in high-confidence computing scenarios \cite{chen2022survey}. Guru et al. (2023) focus on consensus protocols (such as PoW and PoS), analyzing potential attacks targeting these protocols and proposing strategies to enhance system resilience \cite{guru2023survey}. Platt and McBurney (2023) delve into the ability of blockchain consensus mechanisms to resist Sybil attacks, comparing the attack resistance of various algorithms and offering insights for more robust designs \cite{platt2023sybil}. Akbar et al. (2021) innovatively propose a distributed hybrid mechanism that integrates the strengths of PoW and PoS to prevent double-spending attacks, demonstrating its feasibility in future Internet applications \cite{akbar2021distributed}. Chaganti et al. (2022) systematically review DoS attacks within the blockchain ecosystem, highlighting the limitations of existing defense measures and identifying future research challenges \cite{chaganti2022comprehensive}. Meanwhile, Madhushanie et al. (2024) provide a systematic analysis of the mechanisms, impacts, and countermeasures of selfish mining attacks \cite{madhushanie2024selfish}. These studies not only reveal the security threats and technical bottlenecks confronting blockchain technology in its development and application but also offer a solid theoretical foundation and research direction for building more secure and efficient blockchain systems, laying a crucial foundation for the exploration in this paper.

Leveraging the foundational insights from the aforementioned studies, this research undertakes a more thorough and detailed investigation into the security challenges confronting cryptocurrency. Initially, it gathers and organizes a dataset of real-world attack incidents from recent years, systematically filtering them to pinpoint the most prevalent and representative attack types. Next, utilizing the framework of blockchain architecture, it categorizes these attacks into five distinct groups, providing a detailed analysis of their underlying vulnerabilities, the mechanisms of their execution, and approaches to replicating them for study. Lastly, the research assesses the strengths and weaknesses of current attack detection and defense strategies through a comparative analysis, while also discussing potential future research directions to address evolving threats and strengthen the security of cryptocurrency systems.

The main contributions of this study are:
\begin{itemize}
\item \textbf{Comprehensive data collection:} It collects 165 real attack cases and identifies high-frequency attack types through data statistical analysis.
\item \textbf{Complete classification system: }It constructs a systematic security threat and attack classification system based on vulnerability principles.
\item \textbf{In-depth comparative analysis:} It comprehensively summarizes and compares the advantages and disadvantages of existing attack detection and defense schemes.
\end{itemize}
This study aims to provide valuable references for research in the field of cryptocurrency security and promote the innovation and development of related defense technologies.

\section{Background}

\subsection{Related Content}

\subsubsection*{\textbf{BlockChain}}
Blockchain technology is a decentralized data storage and management model. It achieves data immutability, disintermediation, and distributed management by constructing a distributed ledger \cite{li2023overview}. Its core advantage lies in the "decentralization" theory, which ensures data reliability and security through communication and data sharing among multiple nodes in the network, where all nodes have equal status, thus avoiding the risk of single points of failure \cite{bennet2024blockchain}. To ensure data security, blockchain employs technologies such as hash algorithms and asymmetric encryption \cite{wang2024blockchain}.

Bitcoin, the first cryptocurrency based on blockchain, is proposed by Satoshi Nakamoto in 2009 \cite{depoortere2024examining}. Its success promotes the development of the global cryptocurrency market. Bitcoin's blockchain technology verifies and records transactions through a distributed ledger, protects data and value, and maintains market stability. Each transaction is encrypted and linked into an immutable chain \cite{nasir2024securing}.

The blockchain network architecture typically divides into six layers: the core data layer, network layer, and consensus layer, as well as the extension layers of the incentive layer, contract layer, and application layer. The first three are the foundation for implementing blockchain technology.

\begin{figure*}
\centering
\includegraphics[width=\textwidth]{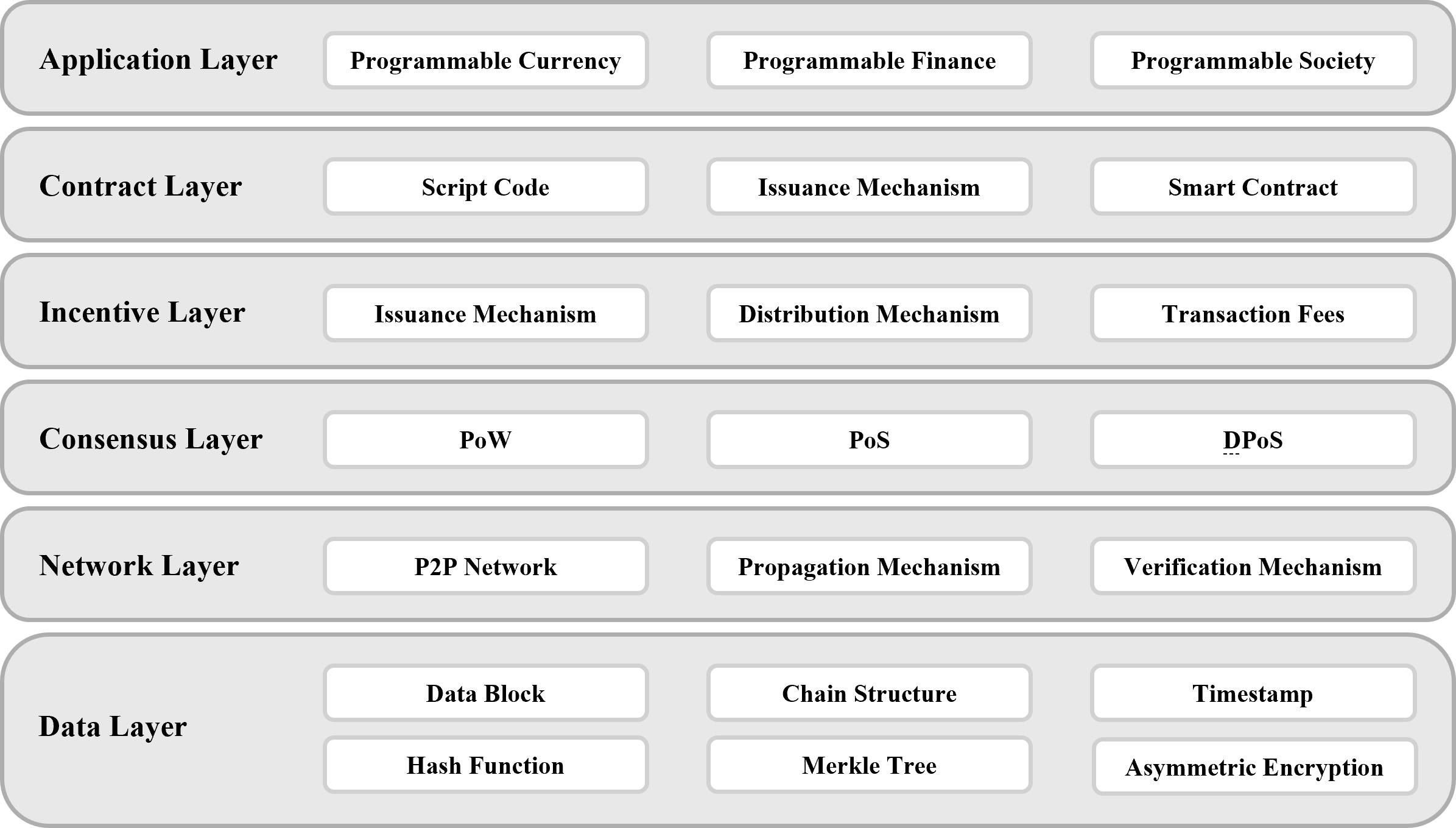}
\caption{The Six Layers of Blockchain Network Architecture}
\label{fig:overraching}
\end{figure*}

\subsubsection*{\textbf{Smart Contracts}}
A smart contract is an automatically executing contract protocol based on blockchain technology \cite{li2024stateguard, li2024scla}. It primarily runs on blockchain systems such as Ethereum. Its purpose is to ensure the security, fairness, and transparency of transactions between parties without the need for intermediaries. Essentially, a smart contract is a piece of computer code that consists of multiple code modules and can implement complex business logic. It has characteristics such as self-execution, immutability, and decentralization. It can be used to create cryptocurrency, manage digital assets, determine voting results, and so on \cite{li2020characterizing}.

\subsubsection*{\textbf{Lightning Network}}

Lightning Network as a layer-two solution to Bitcoin's scalability problem, and it has been developing since then \cite{lisi2021lightnings}. It allows the network to conduct millions of transactions per second, thereby significantly increasing scalability.

For example, if Alice and Bob need to conduct multiple token transactions within the network, recording each transaction on the mainnet creates a significant waiting time until the previous transaction record completes before the next transaction can take place. The Lightning Network, however, proposes a bilateral payment channel that transfers transactions to a side network. If one party wants to stop the transactions, they simply upload the final transaction result to the mainnet.

Another advantage of the Lightning Network is that if Alice and Eric are strangers, Alice has an open payment channel with Bob, Bob has an open payment channel with Dale, and Dale has an open payment channel with Eric. When Alice tries to transact with Eric, she only needs to ensure that Bob and Dale have sufficient transaction funds in their accounts. Alice can then achieve the transaction with Eric through her friend Bob, and Bob's friend Dale.

\subsubsection*{\textbf{Mining Pool}}

Due to the continuously rising overall computing power of the blockchain network, it is difficult to compete for the right to record transactions by relying on the computing power of a single node. Therefore, a mechanism that can combine the computing power of multiple nodes, namely a mining pool, is proposed \cite{wen2022exploration}.

Nodes in a mining pool are divided into administrator nodes and miner nodes. Miner nodes mine in the pool using the mining pool administrator's public key, while the mining pool administrator distributes rewards according to the miners' contribution level.

This mechanism provides corresponding protection for both miner nodes and mining pool administrators. Miner nodes can improve mining efficiency through cooperation, and regardless of whether they mine a complete and valid hash share, they can obtain rewards distributed by the administrator according to their contribution level. Similarly, the hash shares mined by miner nodes are useless to them individually; only through the private key held by the administrator can new blocks be generated, thereby obtaining the block reward \cite{mihaljevic2022approach}.

\subsection{Preliminary Preparation}

\subsubsection*{\textbf{Collection of Attack Events}}

We collect a total of 165 cases of blockchain-based cryptocurrency attacks through channels such as the Ethereum community forum and blockchain research reports.

The dataset composed of these cases includes the most publicly verifiable cryptocurrency attack events. In addition to the DAO event attack, Parity wallet smart contract vulnerability, Coincheck exchange theft, Binance exchange API key theft, and Poly Network cross-chain attack that the introduction mentions, it also includes many other important cases, such as the Bitfinex exchange theft, KuCoin exchange theft, Harvest Finance attack, and 51\% attack events.

This data provides an important reference for research on security threats and attacks related to blockchain-based cryptocurrency. The analysis in this paper is based on this dataset.

\subsubsection*{\textbf{Typical Cryptocurrency Attacks}}

During the preparation phase, we retrieve cryptocurrency attack cases up to April 2023. Although the proportion of various attacks in overall cryptocurrency attack cases changes continuously with time and technological advancements, we can still conduct a preliminary analysis based on recent trends combined with historical data to identify several types of attacks with a relatively high proportion:

\begin{itemize}

\item \textbf{51\% Attack.} The attacker controls more than half of the computing power or stake in the blockchain network, thereby firmly seizing the right to record transactions. Through a 51\% attack, the attacker can tamper with transaction records, prevent honest miner nodes from working, and execute malicious behaviors such as double-spending.

\item \textbf{Reentrancy Attack.} The attacker repeatedly calls the target contract's transfer function through a fallback function, thereby stealing assets within the target contract or disrupting the normal execution of the contract.

\item \textbf{Integer Overflow} Vulnerability. In the unit type, when data exceeds the maximum or minimum value that the type can store, an overflow occurs, causing the maximum value plus any value to become the minimum value, and the minimum value minus any value to become the maximum value. This is divided into two types: integer overflow and integer underflow. Attackers can use the former to reduce their own purchase expenses and the latter to withdraw huge amounts from target nodes.

\item \textbf{Sybil Attack.} This mainly targets authorized consensus mechanisms. Attackers create a large number of fake accounts to increase their influence in the consensus process.

\item \textbf{Eclipse Attack.} The attacker controls the target node's routing table, placing it in a fake network controlled by the attacker, thereby achieving other attack purposes.

\item \textbf{Transaction Manipulation Attacks.} These refer to attacks where the attacker alters already confirmed transaction records, leading to double-spending or asset loss for the target node. The success rate of this type of attack is relatively low, but if the attacker possesses sufficient computing power advantage, it can significantly increase the success rate of the attack.

\end{itemize}
In addition, this paper also studies and analyzes other typical attack methods, such as \textbf{collision attacks}, \textbf{transaction delay attacks}, \textbf{resource sharing vulnerabilities}, \textbf{selfish mining attacks}, and \textbf{block withholding attacks}.

\begin{figure*}
\centering
\includegraphics[width=\textwidth]{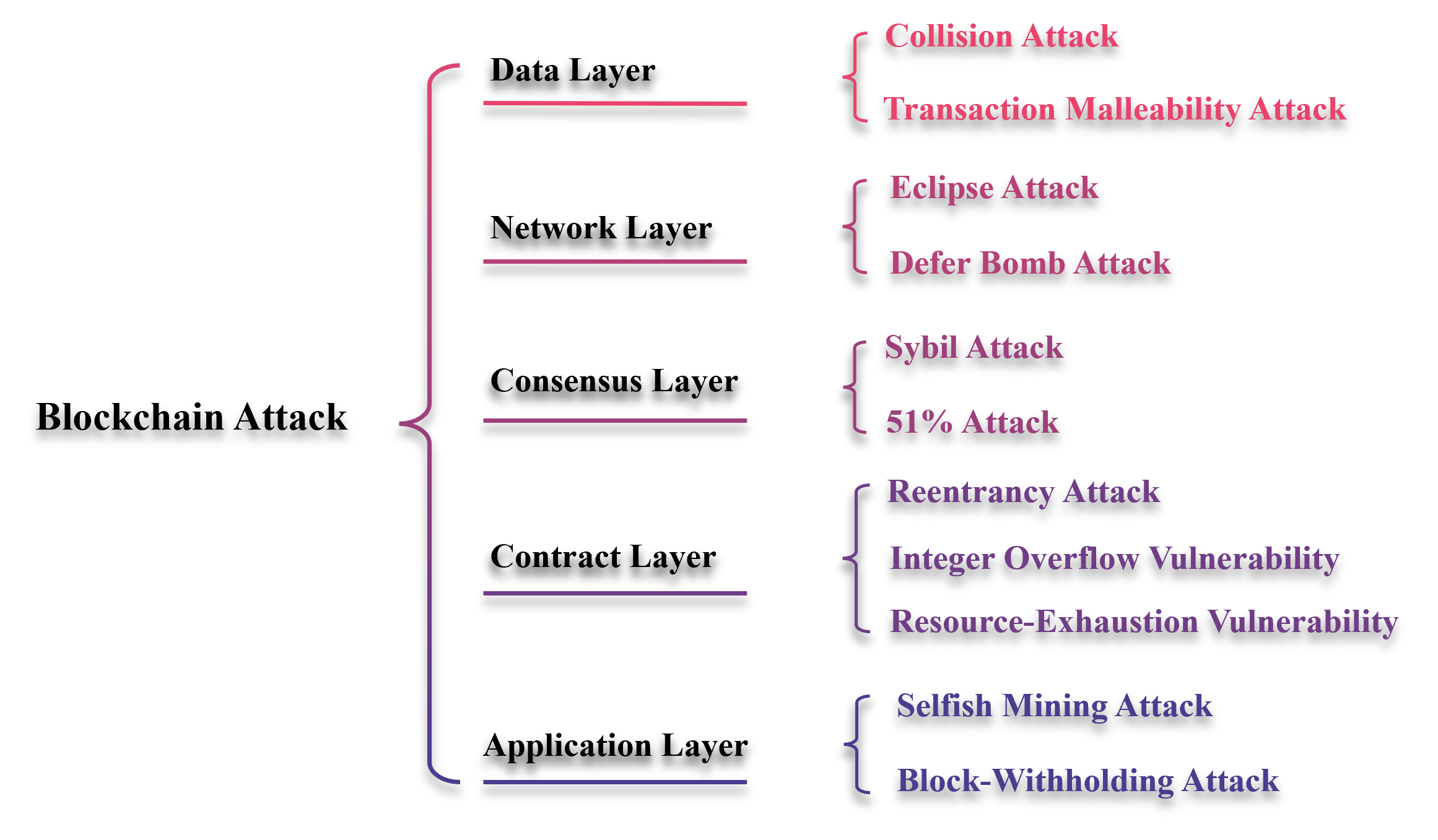}
\caption{Typical Attack Classification}
\label{fig:overraching}
\end{figure*}

\section{Classification Methods}

Leveraging the characteristics of each blockchain layer, we propose a classification method for the security threats and attacks confronting cryptocurrency. This approach categorizes security threats and attacks, analyzing the characteristics, logic, methods, and principles of vulnerabilities exploited by various attack types. During the preliminary preparation phase, we collect a total of 165 attack cases. Through case analysis, we classify typical attacks, and the results are presented in Fig. \ref{fig:overraching}. This classification is informed by research on vulnerability analysis, such as the work on Android applications by Li et al. \cite{li2017discovering}.

\subsection{Data Layer}

Although the data layer features immutability and decentralization, it still faces security threats such as data privacy theft and malicious data attacks \cite{yu2022technology}. Among these, data privacy theft may lead to the leakage of users' personal information, causing losses and risks to users. Meanwhile, malicious data attacks may result in property losses for multiple nodes within the blockchain network, leading to the paralysis of the entire blockchain system and security risks such as data tampering.

\subsubsection{\textbf{Collision Attack}}
A collision attack primarily targets the message digest algorithms in the data layer \cite{kuznetsov2024merkle}. Due to the security guarantees of message digest algorithms, they are hard-coded into the design of blockchain platforms like Ethereum and are regarded as immutable and indestructible building blocks. However, with advancements in cryptanalysis and the epochal increase in computational power, the effective security guarantees provided by message digest algorithms weaken over time.

For message digest algorithms, collisions inevitably exist. Protecting message digest algorithms from collision attacks proves crucial for enhancing the security of the data layer. For example, a victim holds a message $m$ and computes its digest $h(m)$ using a specific message digest algorithm $h(x)$. An attacker only needs to try every possible message $m_i$, compute its digest $h(m_i)$ using the same message digest algorithm, and will certainly find an $m_i$ such that $h(m_i) = h(m)$. From a purely mathematical perspective, in a uniform distribution, the probability of randomly selecting $n$ integers within the numerical range $[1, d]$ and ensuring that at least two of them are identical is given by eq.~\eqref{eq:probability}.

\begin{equation}
P(d, n) = 
\begin{cases} 
1 - \prod_{i=1}^{n-1} \left(1 - \frac{i}{d}\right), & n \leq d \\ 
1, & n > d 
\end{cases}
\label{eq:probability}
\end{equation}

Thus, in practice, such an attack requires enormous computational power, making collision attacks a type of ``future'' attack. Relying solely on computational power makes it difficult to execute a complete attack. Before launching a collision attack, attackers often conduct prolonged cryptanalysis, targeting the most likely datasets for collisions.

\subsubsection{\textbf{Transaction Malleability Attack}}

A transaction malleability attack, also known as a transaction ductility attack or plasticity attack, can be regarded as a variant of the typical double-spend attack in the data layer \cite{zhang2025multi}. It should be noted that this article classifies the typical double-spending attack under the blockchain consensus layer.

\textbf{A. Double Spend Attack} When multiple transaction messages point to the same content, we can determine that these transactions conflict, as only one of them may be valid. An attacker deliberately initiates two conflicting transactions to deceive a third party and gain sufficient benefits. Since blockchain nodes always regard the longest chain as the correct chain, when an attacker issues two transactions for the same funds simultaneously, the blockchain forks and the blockchain defaults to considering the branch containing the first received transaction request as correct, building upon it. However, it also retains the other branch to prevent it from becoming the longest chain.

The attacker exploits this feature to launch a double-spend attack, with the attack process as Fig. \ref{fig:double_spend}:

\begin{figure}[h]
    \centering
    \includegraphics[width=0.8\textwidth]{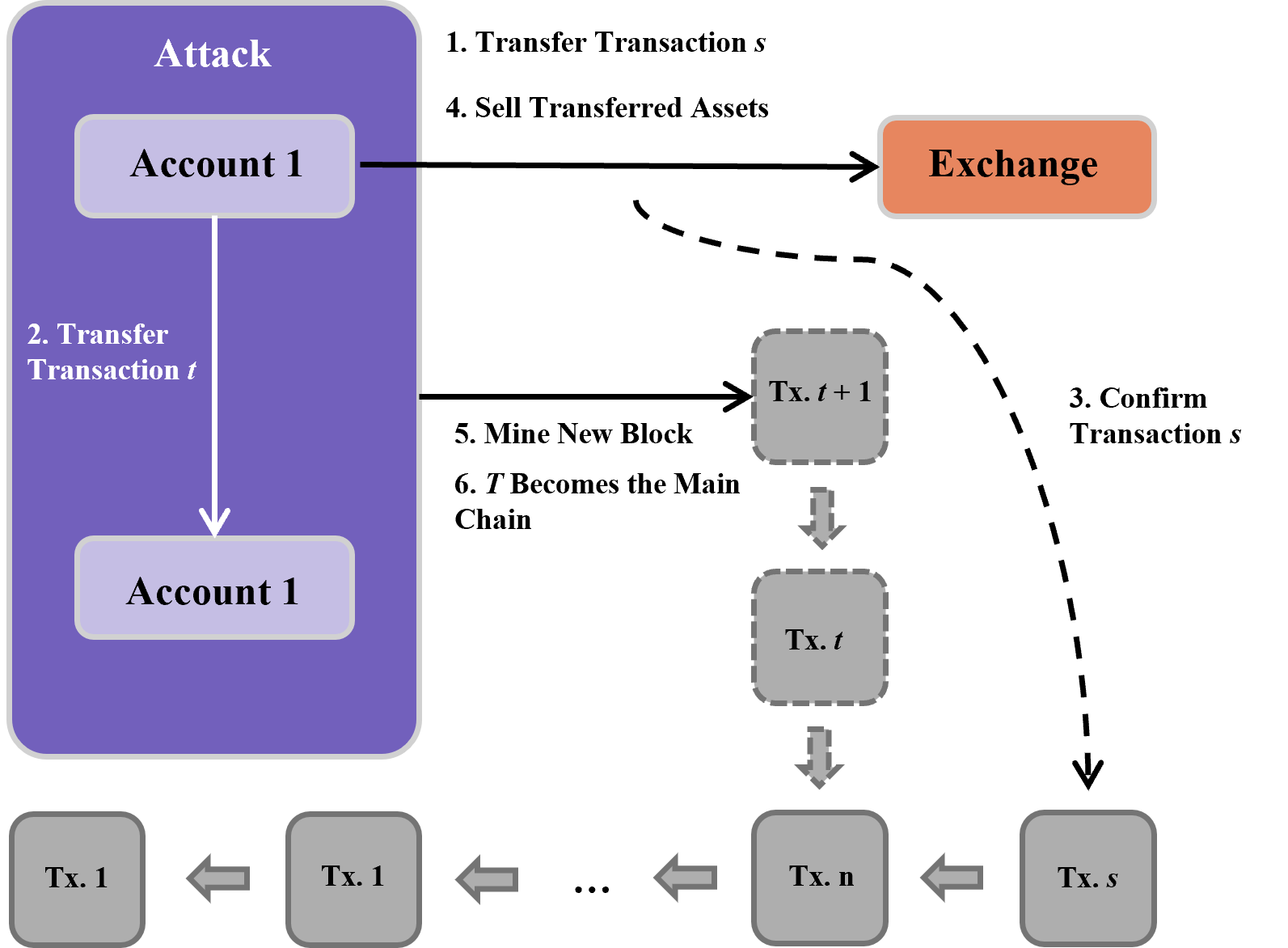}
    \caption{Double Spend Attack}
    \label{fig:double_spend}
\end{figure}

\begin{itemize}
    \item The attacker issues a transaction request $s$, transferring a portion of their funds to a third party that accepts the transaction.
    \item The attacker issues another transaction request $t$, transferring the same funds to another account under their control, causing the blockchain to fork into branch $S$ and branch $T$.
    \item The third party believes they have received the funds from the attacker and confirms transaction request $s$.
    \item The attacker immediately sells these funds, while transaction request $t$ remains unconfirmed.
    \item The attacker mines on branch $T$.
    \item The length of branch $T$ surpasses that of branch $S$, making it the new main chain. Transactions on branch $S$ revert to their previous state, causing the third party to return the funds claimed by transaction $s$ to the attacker, achieving the goal of double-spending.
\end{itemize}

However, executing such a double spend attack proves difficult, as it requires the attacker to control at least 51\% of the computational power to ensure dominance in coin mining, allowing branch $T$ to exceed branch $S$ in length.

\textbf{B. Transaction Malleability Attack}

A transaction malleability attack originates from a vulnerability in the source code of the blockchain system, which allows attackers to alter a transaction’s signature without changing its output or content. Unlike the typical double-spend attack, in a transaction malleability attack, the attacker initiates a transaction not as the sender of funds but as the recipient. The attack process unfolds as  Fig. \ref{fig:transaction_malleability}:

\begin{figure}[h]
    \centering
    \includegraphics[width=0.8\textwidth]{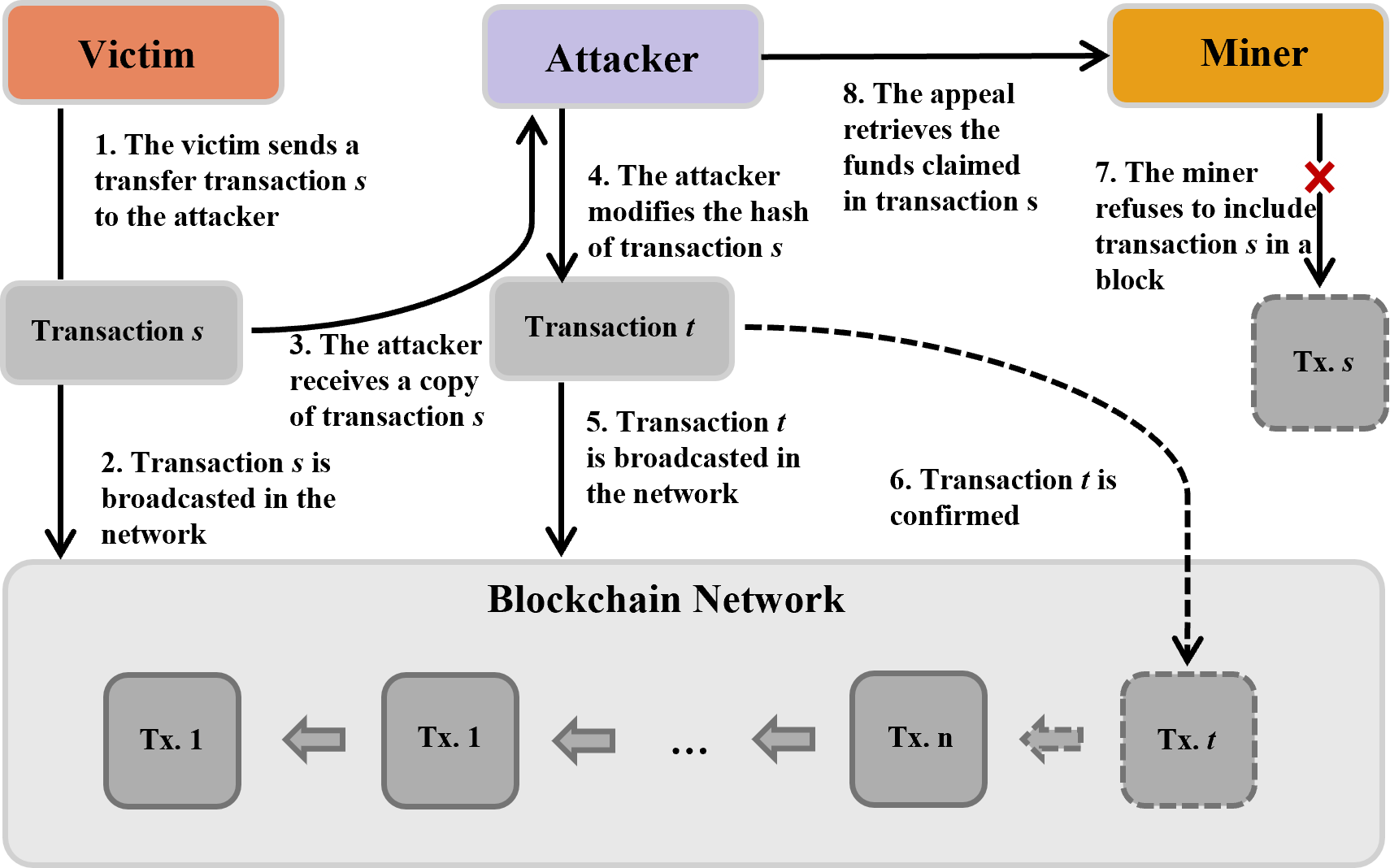}
    \caption{Transaction Malleability Attack}
    \label{fig:transaction_malleability}
\end{figure}

\begin{itemize}
    \item The attacker sends a withdrawal transaction request to the victim, prompting the victim to create a transaction $s$ that transfers a portion of funds to the attacker’s address. At this point, the transaction remains secure and trustworthy.
    \item The attacker waits for the transaction to broadcast across the blockchain network.
    \item The attacker receives a copy of the transaction.
    \item The attacker modifies the transaction information, altering its unique identifier to create transaction $t$.
    \item The attacker waits for transaction $t$ to broadcast across the blockchain network.
    \item Both transaction $s$ and transaction $t$ can be confirmed. If transaction $t$ gets confirmed, the transaction malleability attack succeeds.
    \item If transaction $t$ is confirmed, miners consider transaction $s$ a double spend and refuse to include it in a block.
    \item At this stage, the attack has not yet caused harm to the victim. The attacker needs to file a claim to obtain the funds claimed by transaction $s$. The victim only sees that transaction $s$ remains unconfirmed, while the funds appear credited to the attacker’s account.
\end{itemize}

\subsection{Network Layer}

In the network layer, attackers typically launch malicious attacks targeting the P2P network. The forms of malicious attacks may include delay attacks, node forgery, and more \cite{dai2022ddos}. The network layer also needs to handle issues such as routing selection, congestion control, and internetworking to ensure security and stability. Attackers may exploit these aspects to initiate attacks at the network layer, thereby undermining the security of the blockchain system.

\subsubsection{\textbf{Eclipse Attack}}

An eclipse attack represents a relatively common attack method in the network layer of a blockchain \cite{dai2022eclipse}. Due to the distributed and decentralized nature of blockchain networks, all network nodes can only communicate with each other through adjacent nodes. An attacker modifies the target node’s routing table and adds a sufficient number of malicious nodes around the target node to monopolize all connections with the target node. This isolates the target node from the blockchain network, preventing it from obtaining accurate information. Subsequently, the attacker can exploit the isolated node to perform actions such as forging transactions or executing double-spending attacks, ultimately achieving control over and disruption of the entire blockchain network.

The primary process of an eclipse attack unfolds as follows:
\begin{itemize}
    \item The attacker uses P2P network attacks, such as flooding, to force the victim node to disconnect from its adjacent nodes.
    \item After realizing it cannot communicate with adjacent nodes, the victim may choose to restart, or the attacker may forcibly cause the victim node to restart.
    \item The attacker tampers with the victim node’s routing table, ensuring that, upon restarting, the victim can only connect to malicious nodes controlled by the attacker.
\end{itemize}

\subsubsection{\textbf{Defer Bomb Attack}}

A defer bomb attack, also known as a transaction delay attack, should be distinguished from a transaction malleability attack. A transaction malleability attack represents a variant of a double-spending attack that exploits vulnerabilities in the data layer, whereas a transaction delay attack involves a malicious occupation of network resources.

An attacker launching a transaction delay attack broadcasts a large number of transactions to the network in a short period but does not immediately record these transactions in a block. This causes the confirmation time of these transactions to be relatively delayed, consuming resources from the blockchain network.

The Lightning Network typically employs hash-time-lock technology to ensure the security of atomic asset swaps, with its security primarily relying on time locks and fund locks. A time lock stipulates that each fund swap transaction must be completed within a specific time. However, an attacker can create a large number of fake transactions in a short time and deliberately delay their transmission, leading to network congestion and reduced network performance, thereby affecting the normal operation of the Lightning Network. Therefore, a transaction delay attack is regarded as a congestion attack in the blockchain environment. Although a transaction delay attack alone rarely brings substantial profits to the attacker, it is frequently used as a precursor to other attacks.

\subsection{Consensus Layer}

Consensus layer attacks are divided into attacks targeting authorized consensus mechanisms and attacks targeting unauthorized consensus mechanisms \cite{guru2023survey}. Both types of attacks essentially involve the attacker using certain methods to prevent the entire network of nodes from reaching a correct consensus.

\subsubsection{\textbf{Sybil Attack}}

Douceur first introduces the concept of a Sybil attack in P2P networks \cite{platt2024sybil}. In an authorized consensus mechanism, each node exerts the same influence on the consensus process. Therefore, an attacker can influence the entire network of nodes in reaching a correct consensus by controlling multiple fake accounts or blockchain nodes, thereby achieving the intended attack behavior. The attack process unfolds as follows:

\begin{itemize}
    \item A blockchain network operates in a decentralized and distributed manner, with nodes controlled by different identities distributed almost uniformly around the world.
    \item The attacker creates $t$ fake accounts. Due to the authorized consensus mechanism, each fake account holds the same influence in the consensus process as a legitimate account.
    \item In the blockchain network, every decision is made through a collective vote by all accounts. Assuming the number of legitimate accounts in the network is $s$, the attacker only needs to ensure $t \geq s$ to force the entire network of nodes to reach an incorrect consensus.
\end{itemize}

By launching a Sybil attack, the attacker can control the voting or ranking process, thereby gaining an advantage over legitimate accounts in the competition and undermining the fairness of the blockchain network \cite{li2024detecting}.
\subsubsection{\textbf{51\% Attack}}
A 51\% attack, also known as a majority attack, constitutes an attack targeting unauthorized consensus mechanisms \cite{aponte202151}. This attack method involves an attacker who, by controlling more than 50\% of the entire network’s computational power or ``stake,'' firmly holds the bookkeeping rights in the blockchain network. This allows the attacker to control the entire network and prevent the storage and verification of other blocks. The attacker can then leverage this method to execute other attacks, such as double-spending, selfish mining, and more.

\textbf{A. 51\% Attack in PoW Systems} In a Proof of Work (PoW) system, miners expend computational power to gain bookkeeping rights \cite{hao2022research}. If an attacker or their group controls more than half of the computational power in the blockchain network, they can initiate a 51\% attack. At this point, the attacker possesses the ability to arbitrarily modify transactions and can even use their computational advantage to generate a new branch chain, replacing the current main chain.

After successfully executing a 51\% attack, the attacker can arbitrarily manipulate and modify information on the blockchain. Specifically, the attacker can exclude or alter the order of transaction actions, preventing some or all transactions from being confirmed, and obstructing the normal mining operations of some or all miners. The attacker can also use a 51\% attack as a sub-attack to facilitate the following types of attacks:

\begin{itemize}
    \item \textbf{Double-Spending Attack}: The attacker deliberately initiates two conflicting transactions and uses a 51\% attack to reverse transactions, deceiving a third party to gain sufficient benefits.
    \item \textbf{Transaction Malleability Attack}: Through a 51\% attack, the attacker can effectively ensure that a modified transaction $T$ gets confirmed, thereby guaranteeing the success rate of a transaction malleability attack. However, a 51\% attack is not necessary for a transaction malleability attack.
    \item \textbf{Selfish Mining Attack}: The attacker can leverage their computational advantage to prioritize gaining bookkeeping rights but does not immediately publish the new block. Instead, they continue mining on this block to ensure a sustained advantage in obtaining bookkeeping rewards.
    \item \textbf{History-Revision Attack}: A history-revision attack often occurs after a 51\% attack due to the attacker’s inability to continuously secure bookkeeping rights \cite{rouzbahani2024blockchain}. Honest miner nodes, originally victims, may turn into attackers to minimize their losses by launching a 51\% attack on the new main chain, making their mined branch chain the main chain again. A history-revision attack often becomes a repetitive process, during which the roles of the victim and attacker continuously shift.
\end{itemize}

\textbf{B. 51\% Attack in PoS Systems} In a Proof of Stake (PoS) system, the proportion of assets a node invests in each round of competition for bookkeeping rights is recorded as its ``stake'' \cite{aponte202151}. The more assets invested, the greater the ``stake'', and the higher the advantage in competing for bookkeeping rights. Once the attacker’s invested assets exceed half of the total assets competing for bookkeeping rights in a given round, they gain a significant advantage in securing the bookkeeping rights for that round, thereby achieving control. Unlike in a PoW system, the risks an attacker faces when executing a 51\% attack in a PoS system far exceed the expected gains.

In a PoW system, an attacker launching a 51\% attack incurs no equipment loss, whether the attack succeeds or fails. However, in a PoS system, even if the attack succeeds, the cryptocurrency based on PoS likely depreciates, and the attacker, holding the largest stake, suffers the greatest loss.

In a study by Lee and Kim, a short-selling attack applicable to PoS systems is proposed. Short-selling represents an operation mode in the financial asset domain \cite{cryptoeprint:2020/019}. Suppose Alice borrows an asset $p$ from Bob and sells it at market price to obtain funds $f$. Under stable market conditions, Alice should repurchase asset $p$ with funds $f$ after a period and return it to Bob. However, if the market price of the asset changes during this period, making funds $f$ significantly greater than the current market price of asset $p$, Alice profits from the difference.

In a PoS system, cryptocurrency exchanges exist, making ``short-selling'' theoretically possible:
\begin{enumerate}
    \item The attacker holds a quantity $A$ of tokens, ensuring these tokens are sufficient to launch a 51\% attack, i.e., exceeding half of the total tokens invested in the competition for that round.
    \item The attacker borrows the entire quantity of tokens from an exchange, denoted as $B$.
    \item The attacker expends the full purchasing power of the borrowed tokens to transfer assets, which can be used to purchase tokens in a PoW system for the next ``short-selling attack.'' Let the purchasing power of the tokens at this time be $i$.
    \item The attacker uses malicious methods such as double-spending attacks or block-dropping attacks to degrade the performance of the current blockchain network, causing the tokens to depreciate.
    \item The attacker purchases a quantity $B$ of tokens to return to the exchange, while the purchasing power of the tokens has now decreased to $l$.
    \item The total assets the attacker possesses after completing the ``short-selling attack'' can be calculated as $(i - l)B + lA$, with a final profit of $(i - l)(B - A)$.
\end{enumerate}

\subsection{Contract Layer}

Based on the programming languages and runtime environments of smart contracts, security threats and attacks at the contract layer can be categorized into those targeting smart contracts and those targeting the contract virtual machine \cite{ferreira2021eye, wang2024smart}. The former primarily stem from Solidity language characteristics and developers' non-standard programming practices, such as reentrancy and integer overflow vulnerabilities. The latter arise from unreasonable code applications and designs, including resource exhaustion vulnerabilities and similar issues.

\subsubsection{\textbf{Reentrancy Attack}}

Reentrancy attacks are a prevalent threat in the Ethereum blockchain, tied to vulnerabilities in Solidity, its official smart contract language \cite{alkhalifah2021mechanism}. These attacks exploit flaws in programming logic, leading to substantial losses. They are classified into single-function attacks, where an attacker repeatedly invokes a vulnerable function via a callback loop, and cross-function attacks, where shared data between functions is exploited. Reentrancy vulnerabilities typically arise in the \texttt{withdraw} function due to transferring funds before updating the balance, enabling attackers to repeatedly call it through an external callback \cite{yang2024uncover}.

\begin{lstlisting}[caption={Reentrancy Vulnerability}]
contract Bank {
    mapping(address => uint) public balances;

    function withdraw(uint256 amount) public payable{
        require(balances[msg.sender] >= amount);
        (bool success, ) = msg.sender.call{value: amount}("");
        require(success);
        balances[msg.sender] -= amount;
    }
}
\end{lstlisting}

Below, we simulate the process of a single-function reentrancy attack in the Ethereum blockchain \cite{li2025scalm}:
\begin{enumerate}
    \item The attacker identifies a \texttt{Bank} contract with a reentrancy vulnerability through program analysis.
    \item The attacker deposits a certain amount of tokens into the \texttt{Bank}, recorded as \texttt{balance}. These tokens suffice to call the \texttt{Bank} contract and initiate the attack.
    \item The attacker calls the \texttt{Bank} contract, attempting to withdraw an amount of tokens, denoted as \texttt{amount}, from the \texttt{Bank}.
    \item In the \texttt{withdraw} function of the \texttt{Bank} contract, the contract first checks if the attacker’s \texttt{balance} in the \texttt{Bank} exceeds \texttt{amount}. If it does, the contract sends \texttt{amount} tokens to the attacker’s account and updates the attacker’s \texttt{balance} to \texttt{balance - amount}.
    \item The attacker calls the \texttt{withdraw} function through an \texttt{attack} function. When the \texttt{withdraw} function sends \texttt{amount} tokens to the attacker’s account, it triggers the \texttt{fallback} function written by the attacker.
\end{enumerate}

\begin{lstlisting}[caption={Attack Function in the Attack Contract}]
    function attack(uint amount) public {
        emit Withdraw(address(bank), amount);
        bank.withdraw(amount ether);
    }
}
\end{lstlisting}

\begin{enumerate}
    \setcounter{enumi}{5} 
    \item In the \texttt{fallback} function, the attacker also calls the \texttt{withdraw} function. Since the previous call to the \texttt{withdraw} function has not yet been completed, the attacker's recorded \texttt{ balance} in the \texttt{ bank} contract does not decrease. As a result, the \texttt{Bank} contract continues to send \texttt{amount} tokens to the attacker’s account, triggering the \texttt{fallback} function again.
\end{enumerate}

\begin{lstlisting}[caption={Fallback Function in the Attack Contract}]
    fallback() external payable {
        // Check if the Bank contract has any balance available for withdrawal
        if (address(bank).balance >= amount) {
            bank.withdraw(amount);
        }
    }
}
\end{lstlisting}

\begin{enumerate}
    \setcounter{enumi}{6} 
    \item Consequently, the \texttt{Bank} contract continuously sends tokens to the attacker’s account until the \texttt{Bank}’s balance is depleted or the gas limit is exhausted.
\end{enumerate}

\subsubsection{\textbf{Integer Overflow Vulnerability}}

The principle of an integer overflow and underflow vulnerability operates as follows: In the Ethereum Virtual Machine (EVM), integer types consist of signed integers (\texttt{int}) and unsigned integers (\texttt{uint}). Generally, the \texttt{uint} type sees more frequent use. In the \texttt{uint} type, when an integer exceeds the upper or lower limit of its designated bit size, the EVM defaults to a modulo operation. This causes the maximum value to wrap around to the minimum value when incremented by one, and the minimum value to wrap around to the maximum value when decremented by one. Integer overflow vulnerabilities can be categorized into integer overflow and integer underflow \cite{sun2021mutation}.

\textbf{A. Integer Overflow} For example, with a \texttt{uint256} type variable, Solidity can only handle values less than or equal to the bit size limit. If we add 1 to the maximum value, it exceeds the storage range, causing the data to become 0. In real-world attacks, an attacker can exploit an integer overflow to reduce the cost of purchasing tokens:

\begin{lstlisting}[caption={Integer Overflow Vulnerability}]
contract Bank {
    mapping(address => uint256) public balances;
    uint256 public num;

    function buyTokens(uint256 amount) public payable {
        num = num + amount; // num is uint256 and can overflow
        balances[msg.sender] += amount;
    }
}
\end{lstlisting}

\begin{enumerate}
    \item The attacker identifies a vulnerable \texttt{Bank} contract through program analysis.
\end{enumerate}

\begin{enumerate}
    \setcounter{enumi}{1} 
    \item Since \texttt{num} is declared as a 256-bit unsigned integer variable, the attacker can construct a value that causes an integer overflow.
    \item The attacker spends a minimal fee to acquire a large number of tokens, which they can then sell for profit.
\end{enumerate}

\textbf{B. Integer Underflow} When using the \texttt{uint256} type, subtracting 1 from the minimum value of 0 causes an underflow, making the data wrap around to the maximum value of the current type. An attacker can exploit an integer underflow to drain the victim’s balance at a low cost:
\begin{lstlisting}[caption={Integer Underflow Vulnerability}]
contract Bank {
    mapping(address => uint256) public balances;

    function withdraw(uint256 amount) public {
        balances[msg.sender] = balances[msg.sender] - amount; // Can underflow
        payable(msg.sender).transfer(amount);
    }
}
\end{lstlisting}
\begin{enumerate}
    \item The attacker identifies a vulnerable \texttt{Bank} contract through program analysis.
\end{enumerate}

\begin{enumerate}
    \setcounter{enumi}{1} 
    \item The attacker analyzes the contract vulnerability. Specifically, the vulnerability arises from the fact that \texttt{ balances[msg.sender] - amount} is calculated as a \texttt{uint}, forcing the result to be greater than or equal to zero.
    \item The attacker constructs a very large \texttt{amount}, causing an integer underflow that drains the \texttt{Bank} balance.
\end{enumerate}

\subsubsection{\textbf{Resource-Exhaustion Vulnerability}}

A resource-exhaustion vulnerability differs from reentrancy attacks and integer overflow vulnerabilities, as it represents a typical attack targeting the contract virtual machine. An attacker deploys malicious code on the Ethereum Virtual Machine (EVM) to maliciously consume system storage and computational resources. Therefore, the virtual machine must implement corresponding restriction mechanisms to prevent the abuse of system resources. For instance, Ethereum introduces the concept of Gas to address this issue \cite{liu2024gastrace}.

\begin{itemize}
    \item \textbf{Gas}: Serves as the basic unit to measure the computational resources consumed by a transaction.
    \item \textbf{Gas Price}: Represents the fee (in Ether) required for one unit of Gas.
    \item \textbf{Gas Limit}: Indicates the maximum amount of Gas a transaction sender is willing to pay for the execution of the transaction.
\end{itemize}

The introduction of Gas ensures that an attacker who intends to perform malicious operations on the Ethereum Virtual Machine must incur a high cost, which most attackers prefer to avoid. Consequently, resource-exhaustion vulnerability attacks gradually fade from attention and, in most cases, are only discussed as a typical example of attacks targeting the contract virtual machine.

\subsection{Application Layer}

The application layer, serving as the carrier of blockchain technology, delivers solutions for a wide range of business scenarios \cite{kong2024characterizing, niu2024unveiling, li2024defitail}. It can be broadly categorized into two domains: mining and blockchain transactions. In this paper, we focus primarily on exploring security threats and attacks within the mining scenario, conducting a comparative analysis of the similarities and differences between selfish mining attacks and block-withholding attacks.

\subsubsection{\textbf{Selfish Mining Attack}}
A selfish mining attack occurs when an attacker discovers a block but does not immediately publish it, instead continuing to mine on that block in secret \cite{madhushanie2024selfish}. Once the chain becomes sufficiently long, the attacker publishes all the mined blocks at once, replacing the existing main chain and establishing a new main chain. The previous main chain reverts to its prior state, rendering the blocks mined by honest miners—despite their significant computational resource expenditure—invalid. The specific attack process unfolds as Fig. \ref{fig:selfish_mining_attack}:

\begin{figure}[h]
    \centering
    \includegraphics[width=0.8\textwidth]{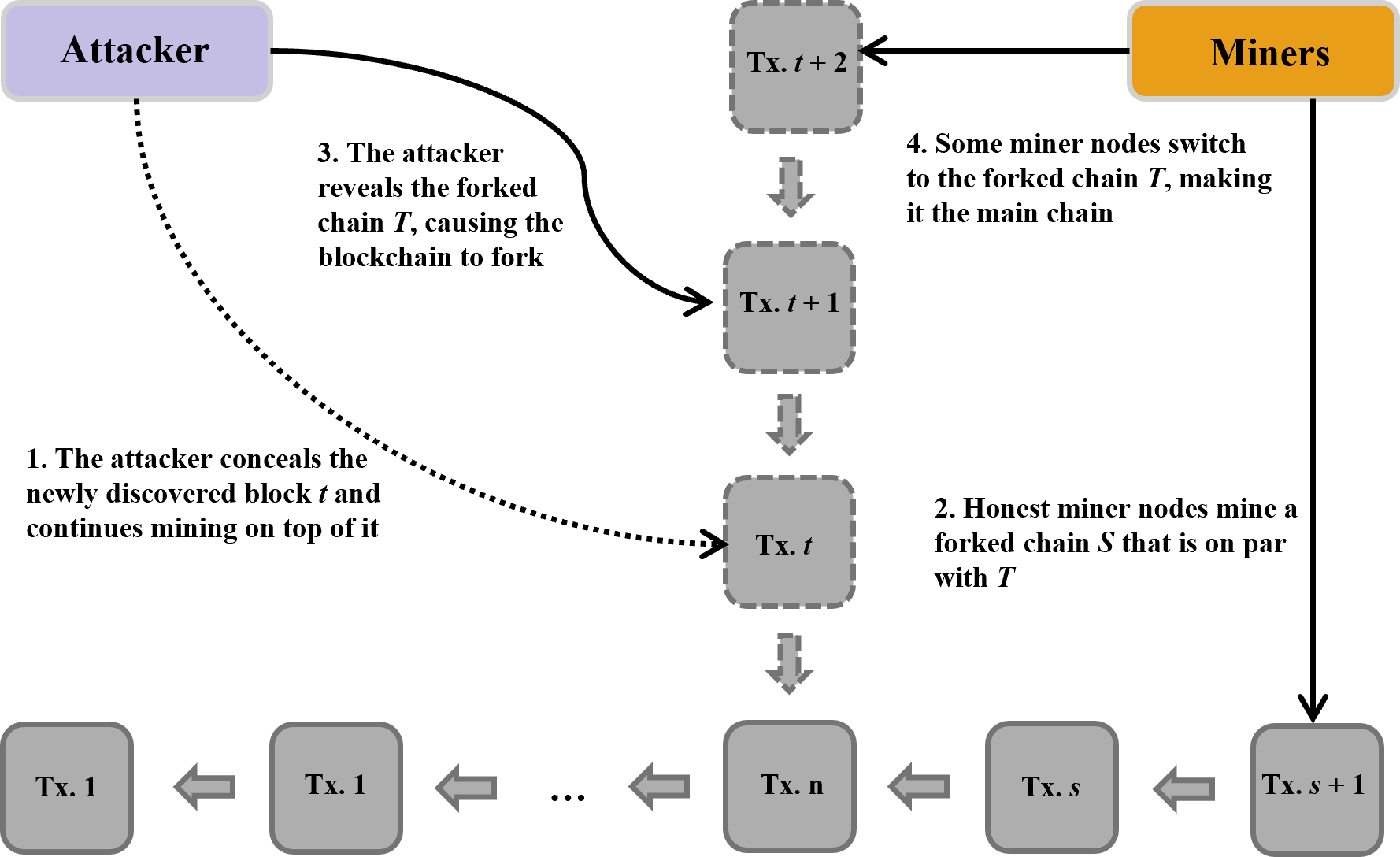}
    \caption{Transaction Malleability Attack}
    \label{fig:selfish_mining_attack}
\end{figure}

\textbf{Assumption 1}: All miner nodes, except the attacker, operate honestly.

Before the selfish mining attack begins, the attacker behaves identically to honest miner nodes. At this point, the blockchain network contains the longest main chain, and all nodes on the chain compete for the next bookkeeping right to earn rewards.

\begin{enumerate}
    \item The attacker discovers a new block first but chooses not to publish it immediately, instead concealing the block or sharing it within an internal network. The attacker continues mining on this block, effectively leading other nodes by one block.
\end{enumerate}

\textbf{Assumption 2}: We consider the worst-case scenario for the attacker, where the blockchain network contains at most two competing branch chains: branch chain $S$, mined by honest nodes, and branch chain $T$, selfishly mined by the attacker. In this scenario, the attacker must compete against the combined computational power of all other nodes.

\begin{enumerate}
    \setcounter{enumi}{1} 
    \item Launching a selfish mining attack requires careful timing for publishing the branch chain. Since honest miner nodes mine simultaneously while the attacker mines new blocks, the attacker does not always have a 100\% guarantee of maintaining an advantage. Two unfavorable situations may arise: a. the length of branch chain $S$, mined by honest nodes, equals the length of branch chain $T$, selfishly mined by the attacker; b. some blocks concealed by the attacker become invalid.
    \item If one of the aforementioned unfavorable situations occurs, the attacker immediately publishes the remaining blocks they hold, releasing branch chain $T$ and causing a blockchain fork.
    \item At this point, the attacker hopes that honest miner nodes switch to mining on chain $T$, allowing chain $T$ to gain a length advantage over chain $S$ in the competition.
\end{enumerate}

When honest miner nodes adopt chain $T$ as the new main chain, the selfish mining attack is considered complete, and branch chain $S$ becomes an invalid branch.

We can also consider an alternative scenario: when branch chain $T$ becomes the main chain, some honest miner nodes may choose to continue mining along branch chain $S$ (especially miners who have published blocks on branch chain $S$). In this case, these honest miner nodes transform into new attackers.

\subsubsection{\textbf{Block-Withholding Attack}}
A block-withholding attack, also known as a block-hiding attack, shares strategic similarities with a selfish mining attack. Proposed by Rosenfeld in 2011, this attack rarely appears in practice \cite{chen2023prevention}. A block withholding attack primarily manifests as an attacker never publishing the blocks they discover, thereby reducing the overall revenue of the mining pool.

Early research categorizes withholding attacks into two types: ``destroy'' and ``wait''. In the former, the attacker does not profit from malicious behavior and is limited to sabotaging the mining pool through malicious means, which incurs significant costs. The latter represents a more complex block-hiding attack similar to selfish mining. Nicolas et al. criticize this attack as impractical.

Nicolas et al. extend the ``destroy'' type attack and propose a scenario where the attacker can profit: the attackers split into two groups, where one group infiltrates other mining pools, claiming rewards while wasting the pool’s computational resources, and the other group mines normally in a private pool, earning bookkeeping rewards. A block withholding attack significantly intensifies malicious competition between mining pools, disrupting the normal mining order. The specific attack process unfolds as Fig. \ref{fig:block-withholding_attack}:

\begin{figure}[h]
    \centering
    \includegraphics[width=0.8\textwidth]{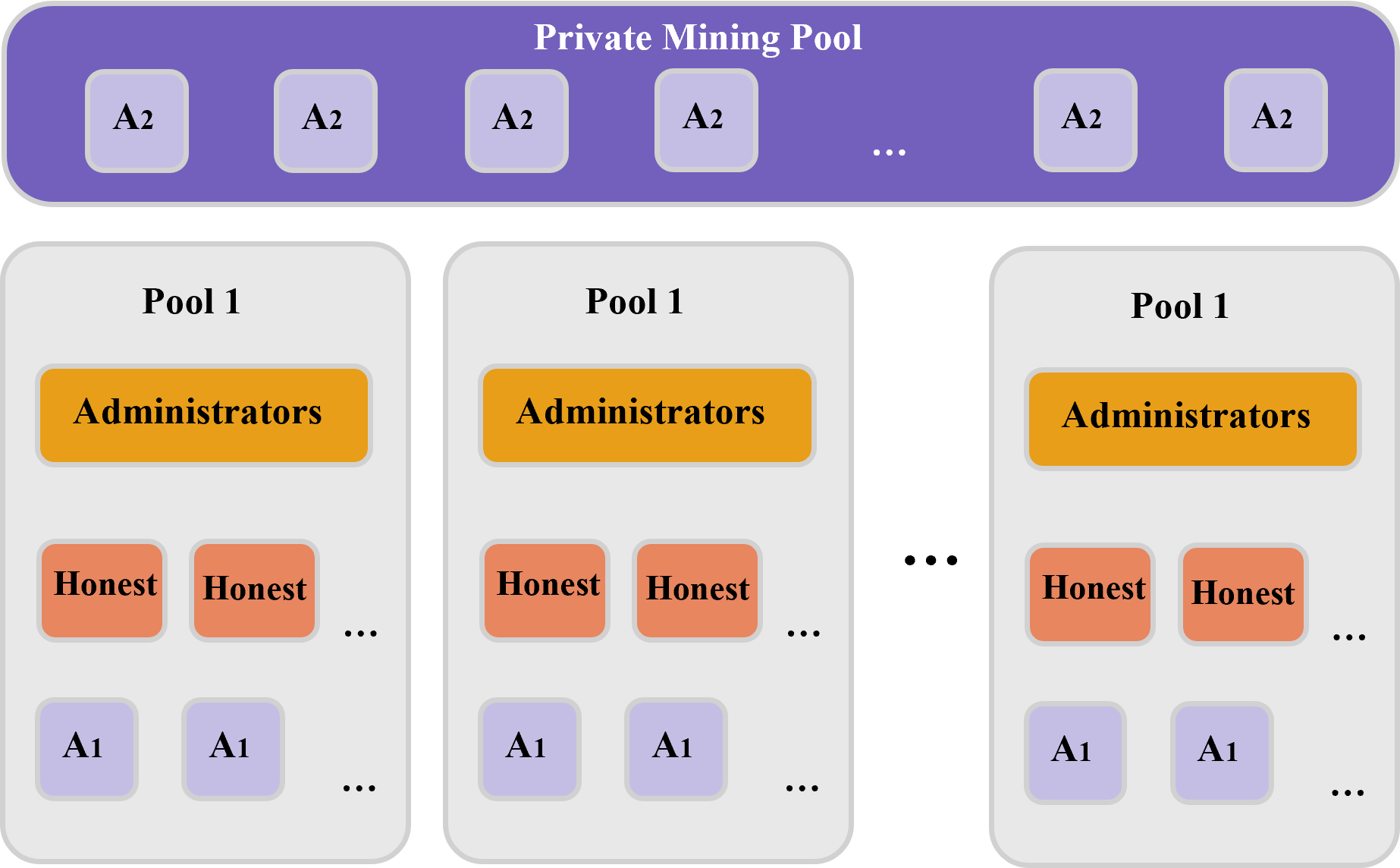}
    \caption{Transaction Malleability Attack}
    \label{fig:block-withholding_attack}
\end{figure}

\textbf{Assumption 1}: All miners use the pool administrator’s public key to mine in the same pool, the pool administrator distributes profits based on miners’ contributions, and the administrator operates honestly.

\textbf{Assumption 2}: Miners do not frequently switch mining pools.

\textbf{Assumption 3}: The computational power of miners in the blockchain network is evenly distributed.

\textbf{Assumption 4}: The attacker’s computational power accounts for $\alpha = 20\%$ of the total computational power of the blockchain network.

\begin{enumerate}
    \item The attackers split into two groups, $A1$ and $A2$, each with computational power of $\frac{\alpha}{2} = 10\%$. Group $A1$ infiltrates the mining pools of the blockchain network in a randomly distributed manner, constantly changing identities to reduce the likelihood of exposure. Including group $A1$, the mining pool collectively controls $1 - \frac{\alpha}{2} = 90\%$ of the total computational power of the blockchain network.
    \item Group $A2$ conducts normal mining activities in a private pool they deploy, which excludes honest miner nodes from joining.
    \item Since the pool administrator distributes rewards based on miners’ contributions, if an honest miner node mines a hash share sufficient to generate a new block, it is immediately sent to the pool administrator to claim a reward. The pool administrator can then use this hash share and their private key to publish a new block, earning a substantial bookkeeping reward.
    \item However, if the hash share is mined by an attacker infiltrating the pool, the attacker conceals it, preventing the pool administrator from earning the bookkeeping reward. Similarly, since the attacker lacks the pool administrator’s private key, they cannot publish the new block either.
    \item The pool administrator cannot determine whether miners engage in block withholding or identify which miner nodes are malicious. Over time, the pool administrator only notices that the pool’s revenue falls far below expectations.
    \item The attackers profit in the following ways: a. Group $A1$ mimics honest miner nodes, sharing rewards while wasting the computational power of the infiltrated pool; b. Group $A2$ mines normally and earns bookkeeping rewards after publishing new blocks.
    \item Since the computational power of the pool infiltrated by group $A1$ accounts for $1 - \frac{\alpha}{2} = 90\%$ of the total computational power of the blockchain network, but group $A1$ conceals the shares they mine, the average revenue of miner nodes in the infiltrated pool decreases by $\frac{1 - \frac{\alpha}{2}}{1 - \alpha} - 1 = \frac{90}{80} - 1 \approx 0.13$.
    \item Meanwhile, group $A2$, mining in the private pool, remains unaffected and gains higher profits: $\frac{1 - \frac{\alpha}{2}}{1 - \alpha} + 1 - 1 = \frac{\alpha}{4(1 - \alpha)} = \frac{20}{4 \times 80} \approx 0.0625$.
    \item Since only group $A2$ can earn bookkeeping rewards, the average profit of the attackers exceeds that of honest miner nodes by approximately 6\% to 7\%, which can be expressed with the eq.~\eqref{eq:expression}.
    
\begin{equation}
2 \times \frac{1 - \frac{\alpha}{2}}{1 - \alpha} + 2 - 1 = \frac{\alpha}{4(1 - \alpha)}
\label{eq:expression}
\end{equation}

    \item Through a block-withholding attack, the attackers achieve profits that exceed their investment, demonstrating that block-withholding attacks are theoretically profitable.
\end{enumerate}

\section{Resolution Strategies}
\subsection{Data Layer}
Most attacks targeting the data layer aim to compromise blockchain data. Therefore, such attacks can be thwarted at their root by adopting advanced cryptographic tools and more secure cryptographic algorithms. Tab. ~\ref{tab:data_layer_defense} presents the defense and detection mechanisms for data layer attacks.

\begin{table}[h]
    \centering
    \caption{Defense and Detection of Data Layer Attacks}
    \begin{tabular}{|m{4.5cm}|p{4cm}|p{4.5cm}|}
        \hline
        \rowcolor{lightgray}
        \textbf{Attack} & \textbf{Defense} & \textbf{Detection} \\
        \hline
        \multirow{2}{*}{Collision Attack} & \multirow{2}{*}{Difficult to detect} & Increase message digest length, introduce timestamp mechanism \\
        \hline
        \multirow{2}{*}{Transaction Malleability Attack} & Verify transaction ID, leverage third-party tools & Use SegWit, multi-signature, transaction ID confirmation \\
        \hline
    \end{tabular}

    \label{tab:data_layer_defense}
\end{table}

\subsubsection{\textbf{Collision Attack}}
Although collision attacks in blockchain networks may still require a long time to emerge on a large scale, preventive measures remain necessary to address potential future attack scenarios.

\begin{itemize}
    \item \textbf{Increase Message Digest Length}: Since message digest algorithms output ciphertexts of the same length regardless of the plaintext’s length, significantly enhancing security becomes possible by increasing the length of the output ciphertext, making attacks more difficult. However, this also increases the computational power required to execute the algorithm, posing another challenge for honest nodes.
    \item \textbf{Introduce Timestamp Mechanism}: Security is achieved by limiting the time available for an attacker to perform a collision attack.
\end{itemize}

\subsubsection{\textbf{Transaction Malleability Attack}}

\paragraph{\textbf{Detection}}
To mitigate transaction malleability attacks, the risk of occurrence can be reduced by verifying the transaction ID before and after broadcasting or confirmation. Furthermore, third-party tools like blockchain.info and TradeBlock can be utilized to monitor the blockchain network for suspicious transaction activities, enabling the timely detection and prevention of transaction malleability attacks by identifying discrepancies in transaction IDs.

\paragraph{\textbf{Defense}}
Core developer Wuille and others propose the Segregated Witness (SegWit) solution to address transaction malleability attacks \cite{basile2022segwit}. The core idea of this solution involves separating the signature information of a block from the transaction data and storing it independently. The hash value generated for the block header depends entirely on the transaction content and the private key held by the node initiating the transaction \cite{kedziora2023analysis}. As a result, an attacker can no longer alter the transaction hash by modifying the signature information, rendering transaction malleability attacks infeasible.

Furthermore, methods such as transaction ID confirmation and multi-signature techniques can also prevent transaction malleability attacks.
\subsection{Network Layer}

The blockchain network layer is primarily susceptible to attacks targeting its P2P network. Attackers frequently exploit the decentralized structure to compromise honest nodes, employing tactics such as routing hijacking, which can intercept or redirect network traffic, and malicious resource occupation, which can overwhelm nodes and hinder their ability to participate. These methods disrupt the blockchain network's operation by impairing communication and consensus processes. Unlike data layer attacks, network layer attacks tend to be more frequent, inflict greater immediate harm by potentially isolating nodes or segments of the network, and have a broader scope of impact on the network's overall availability and stability. Detection and defense mechanisms for these attacks are detailed in Tab. ~\ref{tab:network_layer_defense}.

\begin{table}[h]
    \centering
    \caption{Detection and Defense of Network Layer Attacks}
    \begin{tabular}{|m{4.5cm}|p{4cm}|p{4.5cm}|}
        \hline
        \rowcolor{lightgray}
        \textbf{Attack} & \textbf{Defense} & \textbf{Detection} \\
        \hline
        \multirow{2}{*}{Eclipse Attack} & Leverage node adjacency relationships and information, detect transaction data & enhance node security, increase the number of network nodes, connect only with trusted nodes \\
        \hline
        \multirow{2}{*}{Defer Bomb Attack} & \multirow{2}{*}{Difficult to detect} & Compress attack manipulation time, increase the cost of launching an attack \\
        \hline
    \end{tabular}

    \label{tab:network_layer_defense}
\end{table}

\subsubsection{\textbf{Eclipse Attack}}
\paragraph{\textbf{Detection}}
Alangot et al. propose a method that leverages the adjacency relationships between nodes and the block information of adjacent nodes to detect potential attackers. This method uses the decentralized mechanism of the blockchain to verify detection results, thereby improving the accuracy and reliability of detection \cite{alangot2021decentralized}.

Xu et al. design a detector that analyzes transaction data and node connectivity in the Ethereum network. This detector feeds the collected information into a machine-learning model for training and optimization, enabling the identification of potential eclipse attack behaviors \cite{XU2020101604}.

These two studies provide different approaches and methods for detecting eclipse attacks. From the perspective of experimental data, Alangot et al.’s detection method performs better in terms of efficiency and accuracy, with lower computational and communication overhead compared to other similar methods. Meanwhile, the detector designed by Xu et al. also achieves promising results in experiments and offers scalability and flexibility. However, its experimental sample size remains relatively small, potentially requiring further validation and refinement.

\paragraph{\textbf{Defense}}
To defend against eclipse attacks, strategies such as increasing the number of nodes in the network, enhancing node security, and connecting only with trusted nodes prove effective.

\begin{enumerate}
    \item \textbf{Enhance Node Security}: Providing security support for nodes serves as the most fundamental measure to prevent eclipse attacks. Installing security devices such as firewalls and intrusion detection systems on nodes enables timely detection and blocking of malicious traffic.
    \item \textbf{Increase the Number of Network Nodes}: When malicious nodes launch an eclipse attack, they typically control one or more nodes in the network to manipulate message traffic. Increasing the number of nodes reduces the influence of a single node, making it more difficult for attackers to control the network.
    \item \textbf{Connect Only with Trusted Nodes}: Each node can significantly reduce the risk of connecting to malicious nodes by ensuring that its adjacent nodes are trustworthy, thereby enhancing the overall security of the network. However, this method poses challenges for newly joined nodes. Therefore, a verification mechanism must also be introduced to identify node characteristics, allowing only verified nodes to join the network.
\end{enumerate}

\subsubsection{\textbf{Defer Bomb Attack}}
Currently, several solutions exist to address defer bomb attacks:

\begin{enumerate}
    \item \textbf{Compress Attack Manipulation Time}: Reducing the time threshold for transactions in a pending state and optimizing the transaction processing performance of the blockchain prevent attackers from gaining sufficient time to execute the attack.
    \item \textbf{Increase the Cost of Launching an Attack}: Increasing the cost of launching an attack can be achieved by optimizing transaction fees (Gas) and imposing limits on transaction amounts, making it difficult for attackers to create a large number of transactions at a low cost in a short period.
    \item \textbf{Optimize Transaction Processing Queue}: Randomizing the processing of pending transactions prevents attackers from determining the next transaction to be processed. Alternatively, prioritizing the earliest submitted transactions can prevent attackers from delaying the processing of specific transactions.
\end{enumerate}

However, these solutions are not effective and may even negatively impact the blockchain network. Therefore, when selecting a defense strategy for deferring bomb attacks, it becomes essential to carefully consider various factors, including the current network conditions and specific application scenarios.

\subsection{Consensus Layer}
Attacks targeting the consensus layer can be mitigated by optimizing the consensus mechanism. However, since the behavior of attackers before an attack shows almost no difference from that of honest nodes, detecting such attacks using conventional methods proves challenging. To address consensus layer attacks, a combination of techniques from multiple dimensions—such as data monitoring, behavior analysis, and network traffic monitoring—must be employed to maximize the prevention of their occurrence. Tab. ~\ref{tab:consensus_layer_defense} presents the detection and defense mechanisms for consensus layer attacks.

\begin{table}[h]
    \centering
    \caption{Detection and Defense of Consensus Layer Attacks}
    \begin{tabular}{|m{4.5cm}|p{6.5cm}|}
        \hline
        \rowcolor{lightgray}
        \textbf{Attack} & \textbf{Detection} \\
        \hline
        \multirow{2}{*}{Sybil Attack} & Increase the number of network nodes, introduce identity verification mechanisms, increase the cost of launching an attack\\
        \hline
        \multirow{2}{*}{51\% Attack} & Enhance the advantage of honest nodes, introduce multiple consensus mechanisms \\
        \hline
    \end{tabular}

    \label{tab:consensus_layer_defense}
\end{table}

\subsubsection{\textbf{Sybil Attack}}
The primary strategy for defending against a Sybil attack involves improving the security mechanisms of the blockchain network and enhancing the security and accuracy of the consensus mechanism. Several methods exist to mitigate Sybil attacks, but relying solely on a single method often fails to provide effective defense and may even introduce risks to the overall network \cite{platt2023sybil}.

\begin{enumerate}
    \item \textbf{Increase the Number of Network Nodes}: Attackers executing a Sybil attack typically need to control a large number of fake nodes. Increasing the total number of nodes in the blockchain network reduces the likelihood of a Sybil attack. However, adding a large number of nodes in a short period may impact the blockchain network, so this approach should be combined with other strategies.
    \item \textbf{Introduce Identity Verification Mechanisms}: Implementing multi-factor identity verification for each newly joined node makes it difficult for attackers to create numerous accounts. However, identity verification mechanisms may increase the centralization of the blockchain.
    \item \textbf{Increase the Cost of Launching an Attack}: Adopting a hybrid consensus mechanism can deter Sybil attacks. For example, in a PoW mechanism, requiring nodes to expend real computational power to compete for bookkeeping rights makes the attack costly. Additionally, gradually increasing the cost of creating accounts can meet the needs of some honest nodes for multiple accounts while preventing attackers from controlling a large number of accounts.
\end{enumerate}

\subsubsection{\textbf{51\% Attack}}
A 51\% attack occurs when an attacker controls more than 50\% of the computational power or stake in the entire blockchain network, allowing them to tamper with and control transaction information \cite{aponte202151}. Defending against a 51\% attack often requires a combination of technical and managerial measures to ensure the network’s decentralization, security, and reliability. The following two strategies prove effective:

\begin{enumerate}
    \item \textbf{Enhance the Advantage of Honest Nodes}: In a PoW system, increasing the computational power of honest miner nodes by adopting more advanced equipment can boost the overall computational power of the blockchain network. Encouraging more miner nodes to join the blockchain network also disperses computational power, enhancing the network’s decentralization and reducing the likelihood of an attacker dominating the network’s computational power. In a PoS system, encouraging honest nodes to participate in the competition for bookkeeping rights can reduce the stake advantage of malicious nodes.
    \item \textbf{Introduce Multiple Consensus Mechanisms}: A 51\% attack primarily occurs in blockchain networks that adopt a single consensus mechanism. Introducing different consensus mechanisms such as PoS or DPoS, and using them in combination with PoW, reduces the likelihood of an attacker controlling the computational power or stake, thereby ensuring the network’s decentralization and security.
\end{enumerate}

Additionally, strengthening the management and monitoring of network nodes proves essential to detect abnormal attacker behavior as early as possible. Employing techniques such as multi-signature for nodes and sharding can also enhance the network’s resistance to attacks.

\subsection{Contract Layer}

Attacks on the contract layer typically exploit vulnerabilities in smart contract code or the virtual machine, bypassing the normal logic of the contract to execute an attack. This places higher demands on the security of both the contract code and the contract virtual machine.

Techniques such as code auditing and secure coding practices enable the timely discovery and remediation of vulnerabilities in contracts. Currently, various smart contract detection solutions exist, including static analysis tools and dynamic testing frameworks, which effectively enhance the security and robustness of contracts.

\textbf{Static Analysis}: This involves analyzing the smart contract code or bytecode line by line to identify potential security vulnerabilities \cite{li2024cobra}.

\textbf{Dynamic Analysis}: Dynamic analysis primarily involves detecting smart contracts by simulating their execution, and analyzing potential anomalies that may arise during runtime. Commonly used tools like Echidna and Manticore simulate the execution of smart contracts to identify vulnerabilities within them. Tab. ~\ref{tab:contract_layer_defense} presents the detection and defense mechanisms for contract layer attacks.

\begin{table}[h]
    \centering
    \caption{Detection and Defense of Contract Layer Attacks}
    \begin{tabular}{|m{4.5cm}|p{4cm}|p{4.5cm}|}
        \hline
        \rowcolor{lightgray}
        \textbf{Attack} & \textbf{Defense} & \textbf{Detection} \\
        \hline
        \multirow{2}{*}{Reentrancy Attack} & Static and dynamic detection & code auditing, secure coding techniques \\
        \hline
        \multirow{2}{*}{Integer Overflow Vulnerability} & Static and dynamic detection & code auditing, secure coding techniques \\
        \hline
    \end{tabular}

    \label{tab:contract_layer_defense}
\end{table}

\subsubsection{\textbf{Reentrancy Attack}}
\paragraph{\textbf{Detection}}
In the detector designed by Alkhalifah et al., changes in the state before and after transaction execution, combined with information such as contract addresses and function names, help determine the presence of a reentrancy attack. Predefined security patterns prevent false negatives, while two independent detection mechanisms avoid false positives \cite{alkhalifah2021mechanism}.

Chinen et al. propose a fully static analysis-based detection scheme that evaluates the structure and behavior of the code by fitting an Abstract Syntax Tree (AST) and employing data flow analysis techniques. This approach effectively detects reentrancy attacks in blockchain networks \cite{chinen2021ra}.

Similar to Chinen et al.’s scheme, Yu et al. develop a detection model that uses data flow analysis to identify user-defined attributes and accurately maintain the values of state variables, thereby uncovering reentrancy vulnerabilities in smart contracts \cite{yu2021redetect}.

\begin{enumerate}
    \item Static analysis techniques generate the control flow graph and data dependency graph of the smart contract.
    \item A local graph matching algorithm searches for reentrancy vulnerability patterns in the control flow graph and data dependency graph.
    \item Nodes in the control flow graph and data dependency graph are mapped to a matching template to identify nodes associated with reentrancy vulnerabilities.
\end{enumerate}

Existing detection schemes primarily rely on data flow analysis, static analysis, and dynamic analysis. However, these methods exhibit limitations with external calls and fail to achieve the desired effectiveness in certain application scenarios, making this a key focus for future contract detection schemes.

\paragraph{\textbf{Defense}}
In addition to performing security audits on contract code and employing secure coding techniques, the following methods can be adopted during smart contract deployment to prevent reentrancy attacks:

\begin{enumerate}
    \item \textbf{Use Stricter Programming Statements}: In the Solidity language, contract function calls primarily include \texttt{call()}, \texttt{transfer()}, and \texttt{send()}. Only the \texttt{call()} function triggers the \texttt{fallback} function of an external contract. Therefore, avoiding the use of the \texttt{call()} function can prevent reentrancy attacks. Adding the ``non-reentrant'' keyword to a function ensures that the function completes execution without being repeatedly called. Additionally, setting \texttt{lock(true)} and \texttt{lock(false)} can also prevent a function from being called multiple times.
    \item \textbf{Limit the Number of Transactions with the Same Address}: When the number of transactions with the same account reaches a threshold within a short period, transactions with that account are immediately halted. This functionality, which can be implemented by checking the account address, effectively prevents attackers from performing reentrancy attacks through rapid transactions.
\end{enumerate}

\subsubsection{\textbf{Integer Overflow Vulnerability}}
\paragraph{\textbf{Detection}}
Ayoade et al. employ static analysis techniques to detect vulnerabilities in contract code and create a vulnerability dataset that includes details such as the location of vulnerable statements and operand types \cite{8946210}. By matching arithmetic operations in parts of the contract code with those in the dataset, vulnerabilities are flagged, and the dataset is updated upon detection.

Additionally, Wang et al. design a dynamic analysis-based detection model for unsigned integer overflow vulnerabilities. This model simulates the execution of contract code on the contract virtual machine, checks and constrains the results of unsigned integer operations (such as binary operations, relational operations, and shift operations), and uses the Z3 solver to perform constraint solving, identifying contracts with vulnerabilities \cite{8967006}.

\paragraph{\textbf{Defense}}
Similar to the defense methods for reentrancy attacks, integer overflow vulnerabilities can be mitigated or prevented by using stricter programming statements and limiting external contract calls, thereby eliminating the vulnerabilities or stopping attackers from exploiting them to attack the contract.

\subsection{Application Layer}

Tab. ~\ref{tab:application_layer_defense} presents the detection and defense mechanisms for application layer attacks.

\begin{table}[h]
    \centering
    \caption{Detection and Defense of Application Layer Attacks}
    \begin{tabular}{|m{4.5cm}|p{4cm}|p{4.5cm}|}
        \hline
        \rowcolor{lightgray}
        \textbf{Attack} & \textbf{Defense} & \textbf{Detection} \\
        \hline
        \multirow{2}{*}{Selfish Mining Attack} & Traffic monitoring, establish game-theoretic models & optimize Gas, or introduce other consensus mechanisms \\
        \hline
        \multirow{2}{*}{Block Withholding Attack} & Traffic monitoring, monitor block production time and Bitcoin difficulty adjustment time & introduce penalty mechanisms, or deploy agents to monitor suspicious mining pools \\
        \hline
    \end{tabular}

    \label{tab:application_layer_defense}
\end{table}

\subsubsection{\textbf{Selfish Mining Attack}}
\paragraph{\textbf{Detection}}
Detecting a selfish mining attack requires considering the actual conditions of the blockchain network. The analysis must encompass multiple aspects, including network topology, node transaction behavior, and mining information, to accurately identify attack behavior.

Wang et al. propose a detection scheme based on datasets: by collecting cases of past selfish mining attacks and using analytical techniques to build a detection model, they achieve accurate identification of selfish mining behavior in blockchain networks \cite{wang2021forkdec}. Madhushanie et al., in their study, provide a systematic review of detection approaches, highlighting methods that analyze miner behavior and network patterns to infer selfish mining activities \cite{madhushanie2024selfish}. Additionally, establishing a game-theoretic model between attackers and honest miner nodes allows for the detection of each miner node’s revenue, thereby inferring the presence of attack behavior.

\paragraph{\textbf{Defense}}
Optimizing transaction fees can reduce the occurrence of selfish mining attacks. Research indicates that adjusting economic incentives may deter selfish mining by altering the reward structure, potentially stabilizing blockchain networks \cite{madhushanie2024selfish}. Other researchers also affirm this approach in their articles.

Madhushanie et al. propose a more systematic scheme to defend against selfish mining attacks, detailed as follows \cite{madhushanie2024ba}:

\begin{enumerate}
\item \textbf{Select Appropriate Honest Nodes}: By randomly selecting a sufficient number of active and honest nodes and incentivizing their honest behavior in the blockchain network, these nodes help defend against selfish mining attacks.
\item \textbf{Introduce Multiple Consensus Mechanisms}: Adopting a hybrid consensus mechanism effectively mitigates selfish mining attacks. For example, the Ethereum blockchain introduces the PoS consensus mechanism to reduce the PoW mechanism’s excessive reliance on computational power.
\end{enumerate}

Additionally, schemes such as decentralization, multi-signature mechanisms, and identity verification techniques can also reduce the likelihood of selfish mining attacks.

\subsubsection{\textbf{Block-Withholding Attack}}
\paragraph{\textbf{Detection}}
A block-withholding attack shares certain behavioral similarities with a selfish mining attack, and both can be detected by analyzing network traffic. Furthermore, since block production time and Bitcoin difficulty adjustment time are typically correlated under normal conditions, monitoring these times can also help determine the presence of malicious behavior in the network. Li et al. propose a hybrid statistical test and cross-check method to detect block-withholding attacks, leveraging statistical analysis and multi-node verification for high accuracy \cite{li2024hybrid}.

\paragraph{\textbf{Defense}}
Chen et al. propose a method called the ``forfeiture mechanism,'' which penalizes miners in a pool who maliciously withhold blocks \cite{chen2023prevention}. If a miner is found engaging in malicious behavior, all miners in that pool face penalties. This method fosters closer cooperation between honest miner nodes and the pool administrator, as both aim to identify the attacker as quickly as possible.

Mihaljević et al. propose another defense strategy: deploying a robust consensus protocol into mining pools to counter block-withholding attacks. This approach enhances pool security by optimizing reward distribution and monitoring miner behavior, ensuring the pool’s revenue remains unaffected \cite{mihaljevic2022approach}.

Both defense methods exhibit a degree of innovation but differ in their approaches. Chen et al.’s scheme proves simpler and more direct, causing minimal changes to the blockchain network, while Mihaljević et al.’s scheme appears more advanced and complex.

Additionally, miner nodes can reduce the risk of block-withholding attacks by reasonably distributing computational power and joining fair mining pools.

\section{CONCLUSION}

This paper preliminarily explores the security threats and attacks faced by blockchain-based cryptocurrencies. Based on the characteristics of blockchain layers, these threats and attacks are categorized into five types: data layer attacks, network layer attacks, consensus layer attacks, contract layer attacks, and application layer attacks. This classification facilitates a systematic study of attack characteristics, logic, and processes. Comparative analysis reveals that even within the same layer, different types of attacks exhibit significant differences in principles and implementation methods. For instance, both collision attacks and transaction malleability attacks belong to the data layer, but the former focuses on stealing private data, while the latter targets the blockchain network through malicious data. Furthermore, the study identifies similarities and close relationships between attacks at different layers. For example, a transaction malleability attack represents a variant of a double-spending attack, while a selfish mining attack often requires a 51\% attack as a prerequisite. This paper also conducts a systematic analysis and comparison of existing attack detection and defense methods, evaluating them from multiple perspectives, including security, implementation complexity, and applicability to large-scale networks.

However, the limitations of this study must be acknowledged, particularly in terms of the scope of attack types covered and the depth of the research, both of which require further expansion. Therefore, future research should focus on the following areas to enhance the breadth, depth, and rigor of the study:

\textbf{Expand the Scope of Attack Types:}
Security threats and attacks targeting cryptocurrency exhibit complexity and diversity. This paper only examines relatively typical attack cases, yet attack methods in blockchain networks often lack universality. Future research will aim to expand the dataset of collected attack cases, enabling a broader assessment of potential attack risks.

\textbf{Investigate Composite Attacks and Future Threats:}
As blockchain networks continue to evolve, simple attack methods increasingly struggle to pose significant threats to the system. Attackers now focus more on discovering new vulnerabilities or combining multiple attack techniques. This paper currently lacks sufficient research in this area. Future studies will delve deeper into the characteristics of composite attacks, building on existing cases to explore potential security threats and anticipate future attack methods.

\textbf{Develop More Robust Security Detection Models:}
Current security detection models primarily focus on single attack types, employing static or dynamic detection methods. While these models often achieve expected results in simple case analyses, they struggle to provide accurate feedback in complex, interactive environments. Attackers can exploit various methods to bypass detection mechanisms and successfully execute attacks. Future research will explore the expansion and optimization of detection models to enhance their adaptability and robustness in complex environments.

Through these efforts, this study aims to provide more comprehensive theoretical support and practical guidance for enhancing the security of blockchain networks.

\vspace{10pt}

\bibliographystyle{apalike} 
\bibliography{references} %

\end{document}